\documentclass[sigconf]{acmart}
\usepackage{balance}
\usepackage{multirow}
\usepackage{stfloats}
\usepackage{bm}

\newcommand\blfootnote[1]{%
\begingroup
\renewcommand\thefootnote{}\footnote{#1}%
\addtocounter{footnote}{-1}%
\endgroup
}

\AtBeginDocument{%
  \providecommand\BibTeX{{%
    \normalfont B\kern-0.5em{\scshape i\kern-0.25em b}\kern-0.8em\TeX}}}

\renewcommand\footnotetextcopyrightpermission[1]{} 
\begin{document}
	
%
\title{Human-centric Spatio-Temporal Video Grounding via the Combination of Mutual Matching Network and TubeDETR}

\author{Fan Yu$^{1}$, Zhixiang Zhao$^{1}$, Yuchen Wang$^{1}$, Yi Xu$^{2}$, Tongwei Ren$^{1,*}$, Gangshan Wu$^{1}$}
\affiliation{
\institution{$^{1}$State Key Laboratory for Novel Software Technology, Nanjing University, Nanjing, China}}
\affiliation{
\institution{$^{2}$School of Computer Science and Technology, Soochow University, Suzhou, China}}

\email{{yf, zhaozx, 181250147}@smail.nju.edu.cn, yxu9910@gmail.com, {rentw, gswu}@nju.edu.cn}
	
%
\begin{abstract}
    In this technical report, we represent our solution for the Human-centric Spatio-Temporal Video Grounding (HC-STVG) track of the 4th Person in Context (PIC) workshop and challenge.
    Our solution is built on the basis of TubeDETR and Mutual Matching Network (MMN).
    Specifically, TubeDETR exploits a video-text encoder and a space-time decoder to predict the starting time, the ending time and the tube of the target person.
    MMN detects persons in images, links them as tubes, extracts features of person tubes and the text description, and predicts the similarities between them to choose the most likely person tube as the grounding result.
    Our solution finally finetunes the results by combining the spatio localization of MMN and the temporal localization of TubeDETR.
    In the HC-STVG track of the 4th PIC challenge, our solution achieves the third place.
\end{abstract}

\maketitle
\blfootnote{*Corresponding author.}
\section{Introduction}

Human-centric Spatio-Temporal Video Grounding (HC-STVG) task~\cite{tang2021human} is one of the three tracks in the 4th Person in Context (PIC) workshop and challenge.
HC-STVG is a further exploration of visual grounding, which aims to locate the object of a given query with its bounding box~\cite{hu2016natural,yu2017joint}.
Video grounding requires to localize the starting and ending time of the given video according to a query~\cite{gao2017tall,zeng2020dense}.
Given a sentence depicting an object, spatio-temporal video grounding (STVG)~\cite{zhang2020does,wang2022cross} extracts the spatio-temporal tube of the object.
The query of an input video in HC-STVG is a sentence describing a person in terms of the appearance, the action and the interaction with the environment.
Similar to STVG, HC-STVG needs to localize the target person, \emph{i.e.}, the starting and ending time with the bounding boxes of the target person during the video clip.

The first proposed method for HC-STVG is STGVT~\cite{tang2021human}, which detects region proposals in frames, links the bounding boxes in consecutive frames to form spatio-temporal tube proposals and then uses a visual Transformer combining features extracted from videos and textual descriptions to match and trim the tubes with the given textual description.
Su \emph{et al.}~\cite{su2021stvgbert} propose a unified STVG framework named STVGBert, which also exploits the Transformer model to encode visual and textual features but does not require to generate tube proposals in the begining.
In the 2021 PIC challenge, three more solutions were proposed for HC-STVG.
Tan \emph{et al.}~\cite{tan2021augmented} propose to first localize the temporal segment with the Augmented 2D-TAN model and then predict the spatial location of the target person in each frame.
Yu \emph{et al.}~\cite{chengli2rd} propose to extract human information from the query text, \emph{i.e.,} gender, clothing color and clothing type, generate human tubes from the corresponding video, and finally exploit a Transformer to encode visual and textual features to perform tube-description matching and tube trimming.
Wang \emph{et al.}~\cite{wang2021negative} introduces metric learning~\cite{zheng2021weakly} on the basis of visual features extraction from linked human tubes and textual features extraction from the given query.
Moreover, TubeDETR~\cite{yang2022tubedetr} is proposed as a unified framework for HC-STVG, which uses video-text encoders to combine visual and textual features and predicts starting time, ending time and the spatio-temporal tube with a space-time decoder.

Our solution is built on the basis of TubeDETR~\cite{yang2022tubedetr} and MMN~\cite{wang2021negative}.
We obverse that TubeDETR achieves desired results of spatio localization and MMN has better performance of temporal localization.
Thus, we keep the temporal results of MMN and replace its spatio results with TubeDETR's.

\section{Dataset}

The first dataset for the HC-STVG task is \emph{HC-STVG}, where each video is of 20 seconds and is labeled with a sentence describing a person and the corresponding spatio-temporal localization.
The spatio-temporal localization in \emph{HC-STVG} is represented by the staring frame, the ending frame and the bounding boxes during the segment.
\emph{HC-STVG} dataset has been updated to the third version.
Compared with \emph{HC-STVG} 1.0, data in \emph{HC-STVG} 2.0 are expanded and the labels are cleaned.
In \emph{HC-STVG} 2.1, noisy data are further manually re-annotated and some videos are moved from the test set to the validation set.
The difference among the three versions of data composition is shown in Table~\ref{tab:dataset}.

\begin{table}[htbp]
	\newcommand{\tabincell}[2]{\begin{tabular}{@{}#1@{}}#2\end{tabular}}
	\begin{center}
		\caption{Number of video clips in different versions of \emph{HC-STVG}.}
		\begin{tabular}{c|c|c|c}
			\hline\noalign{\smallskip}
            \multirow{1}{*}{version} & 1.0 & 2.0 & 2.1 \\
			\hline
            \tabincell{c}{training set} &4,500 &10,131 &10,131\\
            \tabincell{c}{validation set} &- &2,000 &3,482 \\
            \tabincell{c}{test set} &1,160 &4,413 &2,913 \\
			\hline
		\end{tabular}
		\label{tab:dataset}
	\end{center}
\end{table}
	
\section{Solution}

\begin{figure*}[htbp]
\centering
\includegraphics[width=\textwidth]{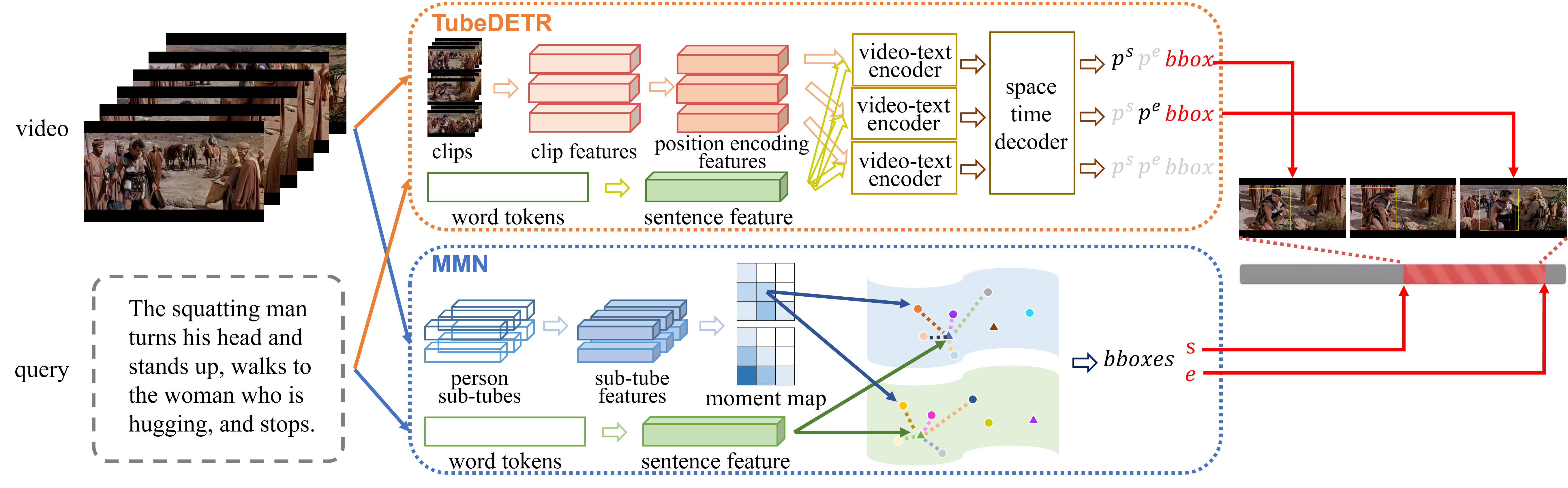}
\caption{Illustration of our solution. $s$ and $e$ represent starting time and ending time respectively, $p_s$ and $p_e$ are probabilities of starting time and ending time respectively, and $bbox$ represents bounding box.}
\label{fig:framework}
\end{figure*}

As illustrated in Figure~\ref{fig:framework}, our solution combines the temporal localization result of MMN and the spatio localization result of TubeDETR.

\textbf{MMN.}
MMN performs cross-modal mutual matching in the metric-learning prospective.
The framework of MMN contains two stages: the first stage aims to extract features and the second stage matches textual description with tube candidates and trims the target tube.
MMN detects humans in frames with Faster R-CNN~\cite{ren2015faster} and links the human bounding boxes following ACT~\cite{kalogeiton2017action} to generate tube candidates.
Each candidate tube is split into 16 clips and each tube clip is considered as a unit.
The visual feature of each unit is generated by CSN~\cite{tan2019lxmert} and a 2D moment map is constructed for each tube candidate to predict the IoU score of a candidate sub-tube for the groundtruth tube with the max-pooled visual features.
To predict the contrastive score, visual feature and textual feature are both used with metric learning.
The final predicted tube is the one containing a moment with the maximum value of the multiplied IoU score and contrastive score, as well as the corresponding starting time and ending time.
During training, the total loss is the summary of iou loss, video loss and sentence loss:
\begin{equation}
	\mathcal{L}_{M} = \mathcal{L}_{M}^{iou}+\lambda(\mathcal{L}_{M}^{vid}+\mathcal{L}_{M}^{sen}),
	\label{con:mmn_loss}
\end{equation}
\begin{equation}
	\mathcal{L}_{M}^{iou} = -\frac{1}{C} \sum^C_{i=1} \left( y_{v_i} \text{log} p_{v_i}^{iou}+(1-y_{v_i})\text{log}(1-p_{v_i}^{iou}) \right),
	\label{con:mmn_iou_loss}
\end{equation}
\begin{equation}
	\mathcal{L}_{M}^{vid} = -\sum^N_{i=1} \text{log} p(i_v|s_i),
	\label{con:mmn_vid_loss}
\end{equation}
\begin{equation}
	\mathcal{L}_{M}^{sen} = -\sum^N_{i=1} \text{log} p(i_s|v_i),
	\label{con:mmn_vid_loss}
\end{equation}
where $\lambda$ is the weight parameter, $C$ is the total number of valid moment candidates, $N$ is the total number of moment-sentence pairs for training, $i_v$ and $i_s$ denote the instance-level classes of the $i^{th}$ moment and the $i^{th}$ sentence respectively, $p_{v_i}^{iou}$ and $y_{v_i}$ denote the predicted and groundtruth iou of the $i^{th}$ moment respectively, and $v_i$ and $s_i$ refer to the $i^{th}$ moment and the $i^{th}$ sentence respectively.

\textbf{TubeDETR.}
Different from MMN, TubeDETR is a unified framework with the encoder-decoder architecture.
The input video is segmented into 20 clips, and the duration of each clips is 1 second.
Visual features extracted from video clips are combined with the textual feature extracted from the corresponding query in video-text encoders.
A space-time decoder then takes the time-sequentially combined features as input and predicts the probability of starting and ending along with the tube for each clip.
During training, the total loss is the summary of bounding box loss, iou loss, Kullback-Leibler divergence loss and guided attention loss:
\begin{equation}
    \begin{aligned}
	\mathcal{L}_{TD}^{sum} =
        \alpha \mathcal{L}_{TD}^{box}+
        \beta \mathcal{L}_{TD}^{Giou}+
        \gamma \mathcal{L}_{TD}^{KL}+
        \theta \mathcal{L}_{TD}^{att},
    \end{aligned}
	\label{con:mmn_loss}
\end{equation}
\begin{equation}
\mathcal{L}_{TD}^{box}=\frac{1}{|B|}\sum_{b \in B}L1(b,\hat{b}),
\end{equation}
\begin{equation}
\mathcal{L}_{TD}^{Giou}=\frac{1}{|B|}\sum_{b \in B}(1-\frac{I}{U}+\frac{A^c-U}{A^c}),
\end{equation}
\begin{equation}
\mathcal{L}_{TD}^{KL}=D_{KL}(\hat{\tau}^s\|\tau^s)+D_{KL}(\hat{\tau}^e\|\tau^e),
\end{equation}
\begin{equation}
\mathcal{L}_{TD}^{att}=-\sum_{i=1}^n(1-\delta_{\tau^s \leq i \leq \tau^e})\text{log}(1-a_i),
\end{equation}
where $\alpha$, $\beta$, $\gamma$, $\theta$ are weight parameters, $B$ is the set of groundtruth bounding boxes, $\hat{b}$ is the predicted bounding box associated with a groundtruth bounding box element $b$, $L1$ represents L1 loss, $I$ and $U$ is the intersection and union area of the predicted bounding box and the groundtruth bounding box respectively, $A^c$ represents the area of the smallest enclosing box, $D_{KL}$ is the Kullback-Leibler divergence, $\hat{\tau}^s$ and $\hat{\tau}^e$ refer to the probabilities of the start and end of the output video tube respectively, $\tau^s$ and $\tau^e$ refer to the target start and end distribution respectively, $\delta$ is the Kronecker delta and $a_i$ is the $i^{th}$ column in the attention matrix $A$.
In our solution, we use the MDETR~\cite{kamath2021mdetr} as the pretrained model, which assists the TubeDETR to achieve the best performance.

\textbf{Finetuning.}
The bounding box results of TubeDETR is directly predicted by the space-time decoder together with the starting time and ending time and the network for jointly spatio-temporal prediction is trained on the HC-STVG dataset.
However, the person tubes and the corresponding features in MMN are generated with pre-trained models.
Thus, the spatio localization of TubeDETR is more accurate than that of MMN.
The temporal location results of MMN are predicted with a starting-ending moment 2D matrix while the starting time and ending time are predicted in TubeDETR independently, Thus, MMN can achieve better performance in temporal localization.
For these reasons, we keep the temporal results of MMN and replace the spatio results with TubeDETR's.
\begin{figure*}[hbp]
\centering
\vspace{-0.1cm}
\includegraphics[width=\textwidth]{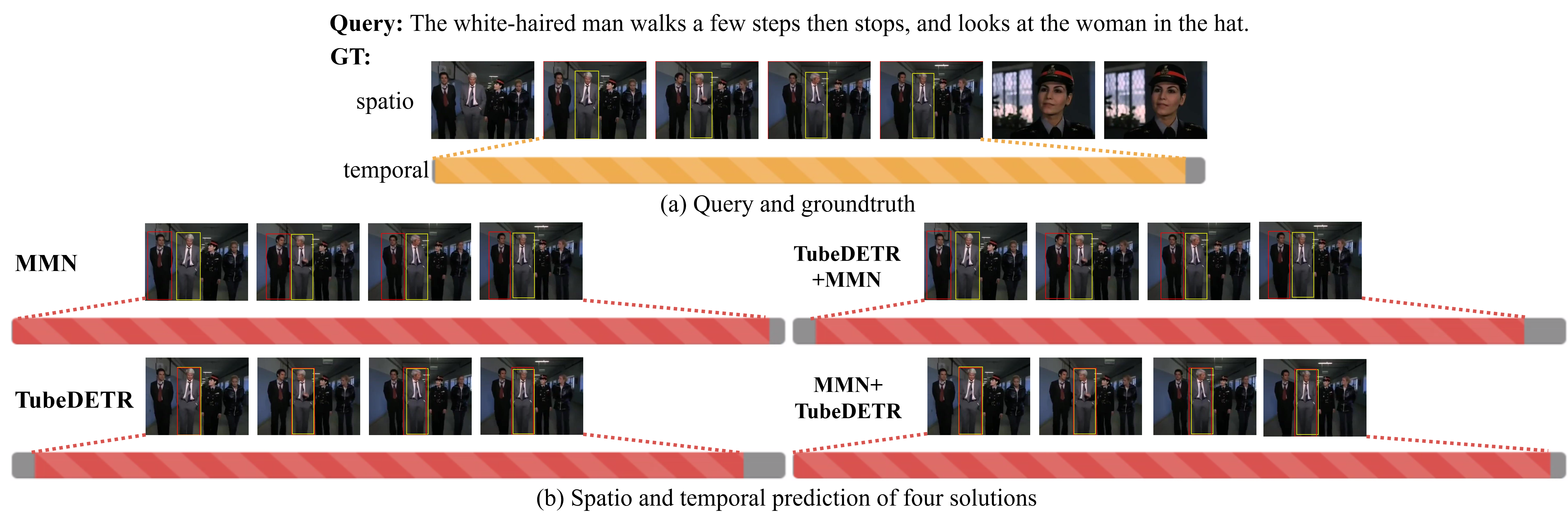}
\caption{An example result of MMN, TubeDETR and MMN+TubeDETR. Spatio and temporal annotations in groundtruth are in yellow, and those in prediction are in red.}
\label{fig:vis1}
\end{figure*}

\section{Experiments}
\subsection{Metrics}

To evaluate the performance of solutions for HC-STVG, three types of metrics are used.

\textbf{tIoU.}
tIoU is used to evaluate the performance of temporal localization:
\begin{equation}
	tIoU = \frac{|S_i|}{|S_u|},
	\label{con:tIoU}
\end{equation}
where $S_i$ is the set of frames in the intersection of predicted and ground truth tube, $S_u$ is the set of frames in the union of predicted and ground truth tube.

\textbf{vIoU.}
vIoU evaluates both temporal localization and spatio trajectory:
\begin{equation}
	vIoU = \frac{1}{|S_u|}\sum_{t \in S_i}IoU(Box^t, Box^{t'}),
	\label{con:vIoU}
\end{equation}
where $Box^t$ and $Box^{t'}$ are the predicted bounding box and ground truth bounding box of frame $t$.

\textbf{vIoU@R.}
vIoU@R represents the percentage of samples whose vIoU is larger than R, and vIoU@0.3 and vIoU@0.5 are used in this report.

\begin{table}[htbp]
	\newcommand{\tabincell}[2]{\begin{tabular}{@{}#1@{}}#2\end{tabular}}
	\begin{center}
		\caption{Comparison results on the \emph{HC-STVG} 2.1 validate set. It is worth noting that the final result of ours in leaderboard of HC-STVG 2022 is the result on the test set of MMN (corresponding to the first line).}
		\begin{tabular}{c|c|c|c|c}
			\hline\noalign{\smallskip}
            \multirow{1}{*}{Methods} & vIoU & tIoU & vIoU@0.3 & vIoU@0.5 \\
			\hline
            \tabincell{c}{MMN} &0.280 &\textbf{0.503} &0.449 &0.227 \\
            \tabincell{c}{TubeDETR} &0.285 &0.445 &0.426 &0.192 \\
            \tabincell{c}{TubeDETR+MMN} &0.255 &0.445 &0.375 &0.154 \\
            \tabincell{c}{MMN+TubeDETR} &\textbf{0.313} &\textbf{0.503} &\textbf{0.501} &\textbf{0.252} \\
			\hline
		\end{tabular}
		\label{tab:quantitative_analysis}
	\end{center}
\vspace{-0.15cm}
\end{table}

\begin{figure*}[htbp]
\centering
\vspace{-0.4cm}
\includegraphics[width=\textwidth]{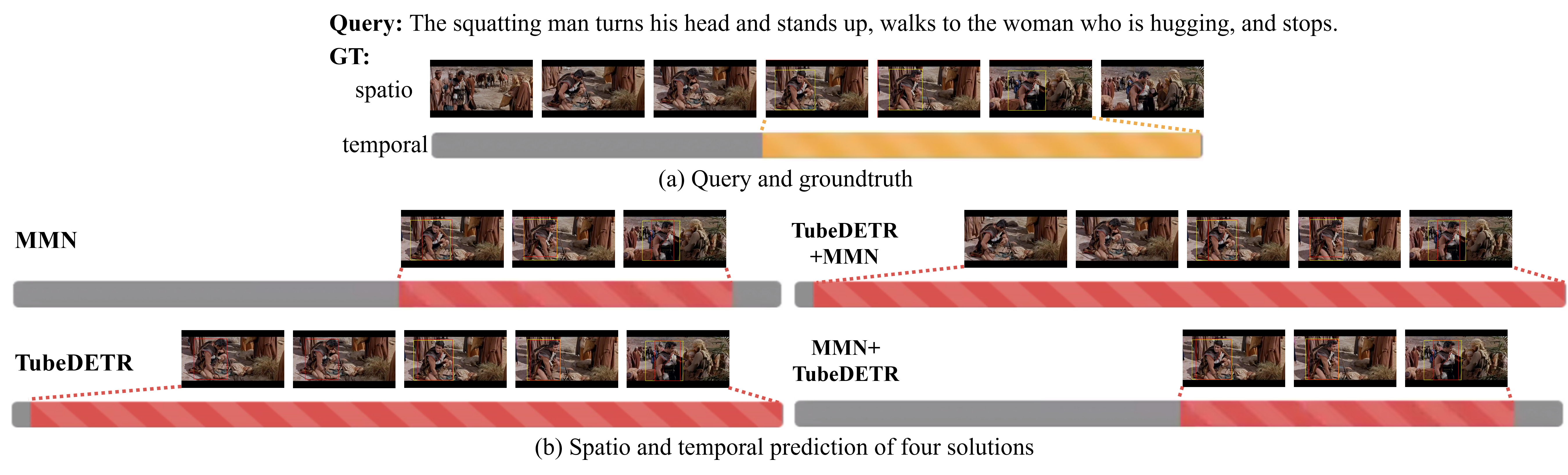}
\caption{Another example result of MMN, TubeDETR and MMN+TubeDETR. Spatio and temporal annotations in groundtruth are in yellow, and those in prediction are in red.}
\label{fig:vis2}
\end{figure*}

\subsection{Quantitative Analysis}
We compare the results of MMN and Tube along with the finetuned results in Table~\ref{tab:quantitative_analysis}.
Compared with MMN, TubeDETR achieves better performance in vIoU but has worse performance in tIoU.
``TubeDETR+MMN'' represents the method that uses the temporal localization of TubeDETR and the spatio localization of MMN, all metrics of which are worse than those of both MMN and TubeDETR.
However, ``MMN+TubeDETR'', which represents the method that uses the temporal result of MMN and replaces its spatio result with TubeDETR's, has the best performance in all metrics.
These experimental data validate the effectiveness of our solution, which combines the temporal localization of MMN and the spatio localization of TubeDETR.

\subsection{Qualitative Analysis}
Two visualization examples (Figure~\ref{fig:vis1} and Figure~\ref{fig:vis2}) show the performance difference between the solutions in Table~\ref{tab:quantitative_analysis}.
As shown in Figure~\ref{fig:vis1}, MMN has accurate temporal localization but detects the wrong person, TubeDETR has accurate spatio localization but its prediction of temporal localization is undesired.
``TubeDETR+MMN'' still detects the wrong person since it keeps the spatio result of MMN, while ``MMN+TubeDETR'' can detect the right person on the basis of the accurate temporal localization.
Figure~\ref{fig:vis2} is another example, where MMN has better performance in temporal localization and TubeDETR almost keeps the whole video duration as the target time, but TubeDETR is more accurate in bounding box detection than MMN.
Since ``TubeDETR+MMN'' uses the temporal result of TubeDETR and the spatio result of MMN, spatio localization is missing in almost half of its target time.
``MMN+TubeDETR'' keeps the accurate temporal localization of MMN and also uses the better spatio localization of TubeDETR, thereby achieving can achieve good performance in both spatio and temporal evaluation.
These examples shows that combining the temporal prediction of MMN and the spatio prediction of TubeDETR is more effective.

\section{Conclusions}
In this report, we represented our solution for the HC-STVG track in PIC 2022 challenge.
Our solution is built on the basis of the MMN and TubeDETR method, keeping the temporal localization result of MMN and the spatio localization result of TubeDETR.
Experiments are conducted on the \emph{HC-STVG} 2.1 dataset and validated the effectiveness of our solution.

\section*{Acknowledgement}
This work is supported by National Science Foundation of China (62072232), Natural Science Foundation of Jiangsu Province (BK20191248) and Collaborative Innovation Center of Novel Software Technology and Industrialization.

\balance
\bibliographystyle{ACM-Reference-Format}
\bibliography{pic22-yuf}

\end{document}